# Flexible Time-Triggered Sampling in Smart Sensor-Based Wireless Control Systems

Feng Xia [1,*] and Wenhong Zhao [2]

1 College of Computer Science and Technology, Zhejiang University, Hangzhou 310027, China
2 Precision Engineering Laboratory, Zhejiang University of Technology, Hangzhou 310014, China

* Author to whom correspondence should be addressed. E-mail: f.xia@ieee.org.

**Abstract:** Wireless control systems (WCSs) often have to operate in dynamic environments where the network traffic load may vary unpredictably over time. The sampling in sensors is conventionally time triggered with fixed periods. In this context, only worse-than-possible quality of control (QoC) can be achieved when the network is underloaded, while overloaded conditions may significantly degrade the QoC, even causing system instability. This is particularly true when the bandwidth of the wireless network is limited and shared by multiple control loops. To address these problems, a flexible time-triggered sampling scheme is presented in this work. Smart sensors are used to facilitate dynamic adjustment of sampling periods, which enhances the flexibility and resource efficiency of the system based on time-triggered sampling. Feedback control technology is exploited for adapting sampling periods in a periodic manner. The deadline miss ratio in each control loop is maintained at/around a desired level, regardless of workload variations. Simulation results show that the proposed sampling scheme is able to deal with dynamic and unpredictable variations in network traffic load. Compared to conventional time-triggered sampling, it leads to much better QoC in WCSs operating in dynamic environments.

# 1. Introduction

In recent years, the great advances in microelectronics and MEMS (micro-electro-mechanical system) have made available inexpensive smart sensors that are equipped with sensing, data processing and wireless communication capabilities. The proliferation of these products in turn makes it possible to design real-time control loops over wireless networks. Consequently, a new generation of networked control systems, i.e., wireless control systems (WCSs), is emerging.

Compared to hard-wired networked control systems, WCSs have many advantages [1-3]. For instance, various difficulties related to the installation and maintenance of the large number of cables normally required are completely eliminated, thus the flexibility and expandability of the system can be further enhanced. At the same time, system maintenance and update become easier, and the cost will of course be reduced. In some harsh industrial environments it is forbidden or unfavorable to use cables due to constraints concerning e.g. physical environments and production conditions. This is especially the case when deleterious chemicals, severe vibrations and high temperatures are present that could potentially damage any sort of cabling. For such situations wireless technologies offer a much better choice for achieving connectivity. In addition, wireless control over sensor/actuator networks satisfies the requirements of mobile systems, enabling closed-loop control of mobile objectives such as automated guided vehicles, mobile robots, and unmanned aerial vehicles.

Some efforts have been made on applying wireless technologies such as Bluetooth (IEEE 802.15.1), WLAN (IEEE 802.11, also called WiFi), and ZigBee (IEEE 802.15.4) to control systems, for example, [4,5]. However, this area is in its infancy at this moment. While most of the work has been done by academic institutions, some commercial companies have also started developing related products for real-world applications. For instance, ABB developed the wireless proximity sensor that offers a solution for addressing reliability and energy conservation issues for fully wireless closed-loop control systems [6]. Meanwhile, the deployment costs of wireless sensor and actuator nodes are continuously decreasing. For example, the node costs for Bluetooth, WLAN, and ZigBee are estimated to be $15, $20, and $10 each, respectively. In contrast, the cost of wiring to connect a node to the existing fieldbus infrastructure is approximately $350 [2], which is obviously much more expensive.

However, the use of wireless networks in connecting spatially distributed sensors, controllers, and actuators raises new challenges for control systems design. Wireless channels have adverse properties such as path loss, multi-path fading, adjacent channel interference, Doppler shifts, and half-duplex operations [1]. Consequently, transmitting radio signals over wireless channels can be affected by many factors, such as ambient noise, physical obstacles, node movement, environmental changes and transmit power, just to mention a few. The inherent openness of wireless connections may potentially cause the operating environment of the system to be highly dynamic, since it is very easy to add new nodes or remove existing ones. Wireless communications are much less dependable than wirelines in that the bit error rate of a wireless channel is usually several times that of a wired connection [7]. Therefore, we will be confronted with more pronounced temporal problems in the form of time-varying delay and packet loss when building control systems upon wireless sensor/actuator networks rather than wirelines [5]. To realize the potential of wireless communication and intelligent sensing technologies in WCSs, new paradigms are required to address these challenges and to provide quality of control (QoC) guarantees in dynamic, unpredictable environments.

This paper is devoted to developing such a paradigm that enables wireless control over smart sensor and actuator networks in the presence of uncertainty in communication resource availability. We are not interested in (robust) controller design or designing novel network protocols for any control systems, which are the topics of most of the work in the emerging field of wireless control. In particular, we will present a flexible time-triggered sampling scheme for smart sensors that are used in WCSs. It attacks the problem of uncertain resource availability deriving from e.g. system reconfiguration and radio interference. The use of smart sensors allows adapting the bandwidth demands of control applications with respect to network conditions through dynamically adjusting the sampling periods. A new algorithm for sampling period adaptation will be developed based on feedback control technology. The QoC of the system that operates in dynamic environments is guaranteed through maintaining the deadline miss ratio (DMR) in each control loop at/around a desired level, since this reduces the impact of both delay and packet loss on QoC.

There has been significant interest in event-triggered sampling that promises to increase resource efficiency, for example, [8-12]. In control applications, however, time-triggered sampling (also called periodic sampling) is still dominant. This is mainly because sampled-data control theory in existence basically originates from time-triggered rather than event-triggered sampling. The theory for event-based control is still under development [13,14]. In systems where the shared computing/communication resource is sufficient, which is usually assumed by control engineers, time-triggered sampling with well-designed fixed periods is able to deliver predictable performance that can be analyzed explicitly using sampled-data control theory. When the resource becomes scarce, however, fixed period based time-triggered sampling will result in worse than possible performance in underloaded conditions and degraded performance even instability in overloaded conditions. There is an obvious lack of flexibility in time-triggered sampling when the system operates in resource-constrained environments with variable workload. The proposed sampling scheme addresses this problem of time-triggered sampling through using flexible sampling periods at runtime. The idea of sampling period adaptation is not new, but we will present a new method for adjusting the sampling period, which improves the flexibility and resource efficiency of the system.

The rest of this paper is organized as follows. In Section 2, we briefly review related work. Section 3 describes the system model to be considered. In Section 4 we present the flexible time-triggered sampling scheme, along with the algorithm for sampling period adaptation. Simulations are conducted in Section 5 to evaluate the performance of the proposed scheme in comparison with the conventional non-adaptive time-triggered sampling scheme. Finally, Section 6 concludes this paper.

## 2. Related Work

Smart sensors (or intelligent sensors) have been applied in various engineering systems, for example, [15,16]. A smart sensor is typically composed of several modules, such as sensing unit, AD (Analog to Digital) converter, microcontroller, storage, transceiver, and power unit. Its capability of data processing enables diverse sampling patterns besides the conventional uniform sampling mechanism. For instance, the concept of send-on-delta, a signal-dependent sampling scheme, has been explored in [10-12] to reduce the number of sensor data transmission. Willett *et al.* [17] proposed an adaptive sampling scheme for wireless sensor networks, which can significantly reduce communications and hence energy consumptions while maintaining high accuracy. Almost all of these works have been done for general-purpose signal processing and telecommunication systems in which no control applications are involved.

In the literature, relatively little progress has been made on applying event-triggered sampling in control systems. Through analyzing first-order stochastic systems, Astrom and Bernhardsson [8] argued that compared to periodic sampling, event-based sampling may require only a fraction of the computing/communication resources while achieving the same control performance. Otanez *et al.* [9] proposed a deadband-based data transmission scheme to reduce network traffic in networked control systems. Nguyen and Suh [18] applied the send-on-delta data transmission method in networked control systems achieving improved estimation performance. Despite the increasing interest in this direction, the lack of a unified theory supporting event-based control has been blocking the practical applications of event-triggered sampling methods in control systems.

Recently, significant work has been done with sampling period adaptation in resource-constrained real-time control systems that basically use time-triggered sampling. For instance, Cervin *et al.* [19] proposed a feedback-feedforward scheduling scheme to dynamically adjust sampling periods that results in near-optimal control performance. Marti *et al.* [20] developed an optimal resource allocation policy that allocates CPU resource in accordance with the current states of controlled systems. In our previous work [21-23], neural network based and fuzzy logic control based feedback scheduling methods for sampling period adaptation in multitasking control systems have been explored, respectively. An overview of this direction can be found in [24]. Since the majority of these papers consider CPU resource constraints and are based on utilization control, the relevant methods are generally not suitable for WCSs where the communication resource rather than the computing resource is the major concern and utilization control could potentially be inefficient.

Li and Chow [25] proposed an adaptive multiple sampling rate scheduling algorithm for Internet-based supervisory control systems. Ploplys *et al.* [7] proposed a method using the PI (proportional-integral) control algorithm to adjust the sampling periods of WCSs over WLAN. Based on the same control structure, Kawka and Alleyne [26] developed another heuristic algorithm to adapt sampling period with respect to packet loss. Colandairaj *et al.* [27] proposed to adapt the sampling periods in response to variations in delay in WCSs, also in a heuristic manner. However, none of these papers considers controlling the deadline miss ratio as we do. Consequently, we can address simultaneously the problems of delay and packet loss, whereas almost all existing methods are dedicated to either of them. In our previous work [28], we have developed a feedback scheduling method to rescale sampling periods based on deadline miss ratio control for multi-loop networked control systems using priority-based fieldbuses. In contrast, this paper focuses on adaptive sampling in smart sensors used in WCSs. The flexible time-triggered concept has been explored in network protocols such as FTT-CAN [29] and FTT-Ethernet [30], whereas it is used for sampling in this paper.

## 3. Wireless Control System Using Smart Sensors

Consider a wireless control system as shown in Figure 1. In addition to some disturbing/interfering nodes (i.e., non-control nodes), there are $N$ independent control loops. For simplicity, assume each control loop is composed of one smart sensor, one controller, and one smart actuator. The sensor and the actuator are attached to the controlled process, which is a single-input single-output (SISO) physical system. All these nodes reside within a *collision area* in which every pair of nodes can *hear* from each other, i.e., all nodes share the same wireless channel. The wireless technology used in the network is ZigBee [31]. Based on the IEEE 802.15.4 specification, ZigBee provides a low-cost and low-power wireless communication solution geared towards wireless sensors and control systems. It fulfils well the unique requirements of applications in which nodes transmit only small data packets and do not require high bandwidth, such as manufacturing automation, process control, home automation, and intelligent building. In the MAC layer, ZigBee uses the CSMA/CA (carrier sense multiple access with collision avoidance) protocol.

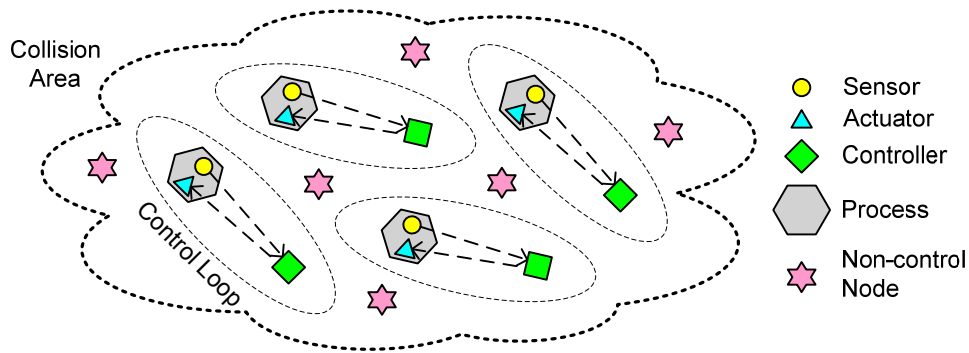

**Figure 1.** Network topology of multiple control loops within a collision area.

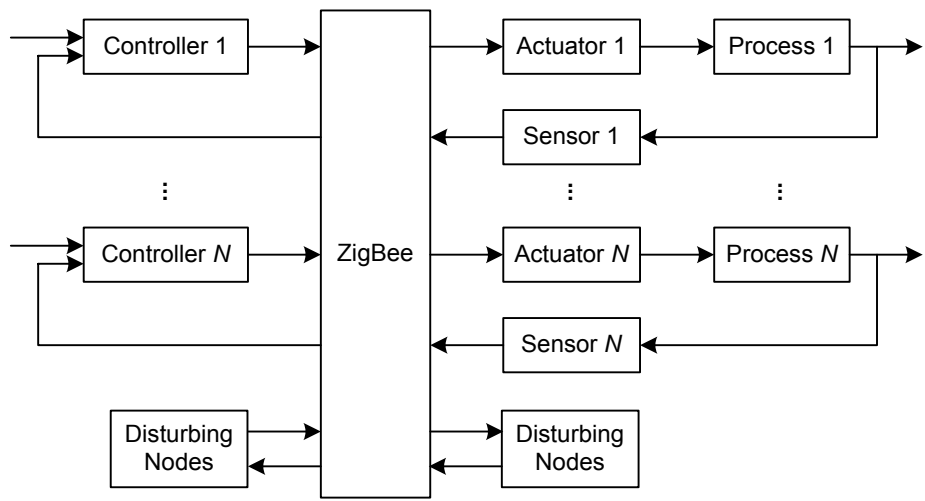

**Figure 2.** Block diagram of the wireless control system.

The block diagram of the WCS is given in Figure 2. In each control loop, the sensor and the actuator communicate with the controller wirelessly. The sensors are time triggered, while the controllers and the actuators are event triggered. Smart sensors and actuators are used to facilitate sampling period adaptation. At the beginning of a sampling period, the sensor collects a measurement of the output of physical process, and then transmits it to the controller via ZigBee. During this term, it may need to compete with other coexisting nodes for the use of the network resource. Upon receiving the sampled data, the controller starts to execute the control algorithm immediately. After the control command is produced, the controller will transmit it, again over the ZigBee network, to the actuator. The actuator will perform the corresponding actions on the physical process once it receives the control command. It is assumed that: 1) the actuator can send data to the sensor belonging to the same loop directly, 2) the sensor and actuator are synchronized in time, and 3) each sampled data and each control command are transmitted over ZigBee as one single data packet, respectively.

Generally speaking, the delay in a networked control loop encompasses sensor processing delay, sensor to controller communication delay, controller computational delay, controller to actuator communication delay, and actuator processing delay. In the above system, the network bandwidth of ZigBee is inherently limited (up to 250Kbps), while being shared by multiple control loops. Consequently, the communication delays are responsible for the largest fraction of the total round-trip delay. Since wireless communication is non-deterministic and CSMA/CA is not a non-destructive protocol, data packets may possibly be lost, due to e.g. too many retransmissions, transmission error and low signal strength. It is well-known in the control community that both delay and packet loss degrade QoC, particularly when they are time-varying.

Intuitively, the delay, packet loss rate, and jitter increase as the network traffic load increases. This is mainly because more collisions on medium access lead to larger waiting delays and larger probability of packets being discarded in sensors and controllers that need to transmit data packets. Therefore, when new control loops are added or interfering radios become present, which causes the traffic load to increase, the QoC of the system may be jeopardized. This is particularly the case if the network is in or close to overload conditions.

Conventionally, the sensors are assigned constant sampling periods. A consequence is that the bandwidth requirements of control loops are also constant at runtime, given that the sizes of the packets to transmit are fixed. When the system operates in dynamic environments with variable workload, the network may be sometimes overloaded, and sometimes underloaded. In overloaded conditions, the QoC may be degraded, as pointed out previously; the system may even become unstable in extreme cases due to severe delay and packet loss problems. In underloaded conditions, on the other hand, the resulting performance may be worse than possible because of low resource utilization. Therefore, new design methods need to be developed to cope with workload variations and to improve the flexibility and resource-efficiency of the system.

In this paper, special attention is paid to the impact of workload variations, which are caused by e.g. system reconfiguration and radio interference, on QoC. To address both delay and packet loss problems that may be induced by workload variations, a flexible time-triggered sampling scheme facilitated by the use of smart sensors will be presented in the next section.

## 4. Flexible Time-Triggered Sampling

By definition, the bandwidth demand of a control loop is closely related to the size of the data packets and the sampling period. With given data packet sizes, the requested network utilization (i.e. workload) of control loops directly depends on the sampling periods of the sensors. This implies that it is possible to regulate the bandwidth demand of a control loop through adjusting the relevant sampling period. From this insight, we propose to adapt the sampling periods of sensors to workload variations at runtime.

The basic idea of our scheme is to maintain the deadline miss ratio in every control loop at a desired level through periodically adjusting the sampling period. Since the sampling period adaptation algorithm will be implemented in every sensor separately, i.e., control loops are independent of each other, we will describe our scheme in this section by considering only one control loop, say loop $i$, and omit this subscript for all variables wherever possible. Despite this, the sampling periods of all sensors will be changed simultaneously at runtime, with the same time interval. To avoid confusion, we call hereafter the periods of control loops (or sensors) *sampling periods* (denoted $h$), while the time interval for executing period adaptation algorithms *invocation interval* (denoted $T_{SPA}$). In this context, the sampling period of each sensor will be re-assigned every $T_{SPA}$ time units with respect to current deadline miss ratio in the corresponding control loop.

A deadline miss occurs when the actuator does not receive the control command by the deadline, which is by default equal to the sampling period. In WCSs, there are generally two types of deadline miss [3]. The first type is that the sampled data or the control command is truly lost in the transmission process. As a consequence, the control command will never arrive at the actuator. In contrast, in the second type of deadline miss, the control command is actually received by the actuator, but at a time point that exceeds the deadline.

In this paper, feedback control technology is used to determine the new sampling period. In control terms, the *controlled variable* is deadline miss ratio, and the *manipulated variable* is sampling period. The deadline miss ratio is defined as the number of deadline misses to the number of periods that the control loop has experienced within a certain invocation interval. Some reasons for the choice of deadline miss ratio as the controlled variable are explained as follows. As one of the most common metrics for network quality of service (QoS), particularly from a real-time viewpoint, deadline miss ratio is an important factor that also affects QoC. Satisfactory QoC can be achieved as long as the deadline miss ratio is controlled at a sufficiently low level. Further, using deadline miss ratio as the controlled variable can address simultaneously the problems of time-varying delay and packet loss. According to the definition of deadline miss, both delays larger than the period and packet losses naturally incur deadline misses. When the deadline miss ratio keeps at a low level, the delays within most sampling periods will be no more than one period and the number of packets being lost is certainly limited. As a consequence, the impact of delay and packet loss on QoC is alleviated.

The sampling period affects the deadline miss ratio in the following way. Shortening sampling period leads to increase in requested network utilization, which naturally causes the network workload to increase, and vice versa. With heavier network traffic load, the probability of collisions between different nodes becomes bigger. This could potentially increase both delay and packet loss, and hence the deadline miss ratio. Therefore, large deadline miss ratio can generally be reduced through enlarging the sampling period, particularly when the system is in overload conditions. When the network is underutilized, on the other hand, the network utilization can be increased through shortening the sampling period. According to sampled data control theory, smaller sampling periods normally deliver better QoC. In this context, the QoC can therefore be improved with higher resource efficiency, given that the deadline miss ratio is limited within a low level. The observation is that, by means of sampling period adjustment with respect to network condition, dynamic and unpredictable workload variations in the WCS can be dealt with effectively. This explains why sampling period is chosen as the manipulated variable.

Figure 3 shows the flexible time-triggered sampling scheme proposed in this paper. Just as the name implies, this scheme is based on time-triggered sampling. Basically, the sensor samples the system output at uniform time intervals. The major difference between our scheme and conventional time-triggered scheme is that the sampling period will be changed regularly with our scheme, whereas the conventional scheme normally uses fixed sampling periods. This results in significantly enhanced flexibility of the system, and hence largely improved QoC in dynamic environments, as will be shown in Section 5.

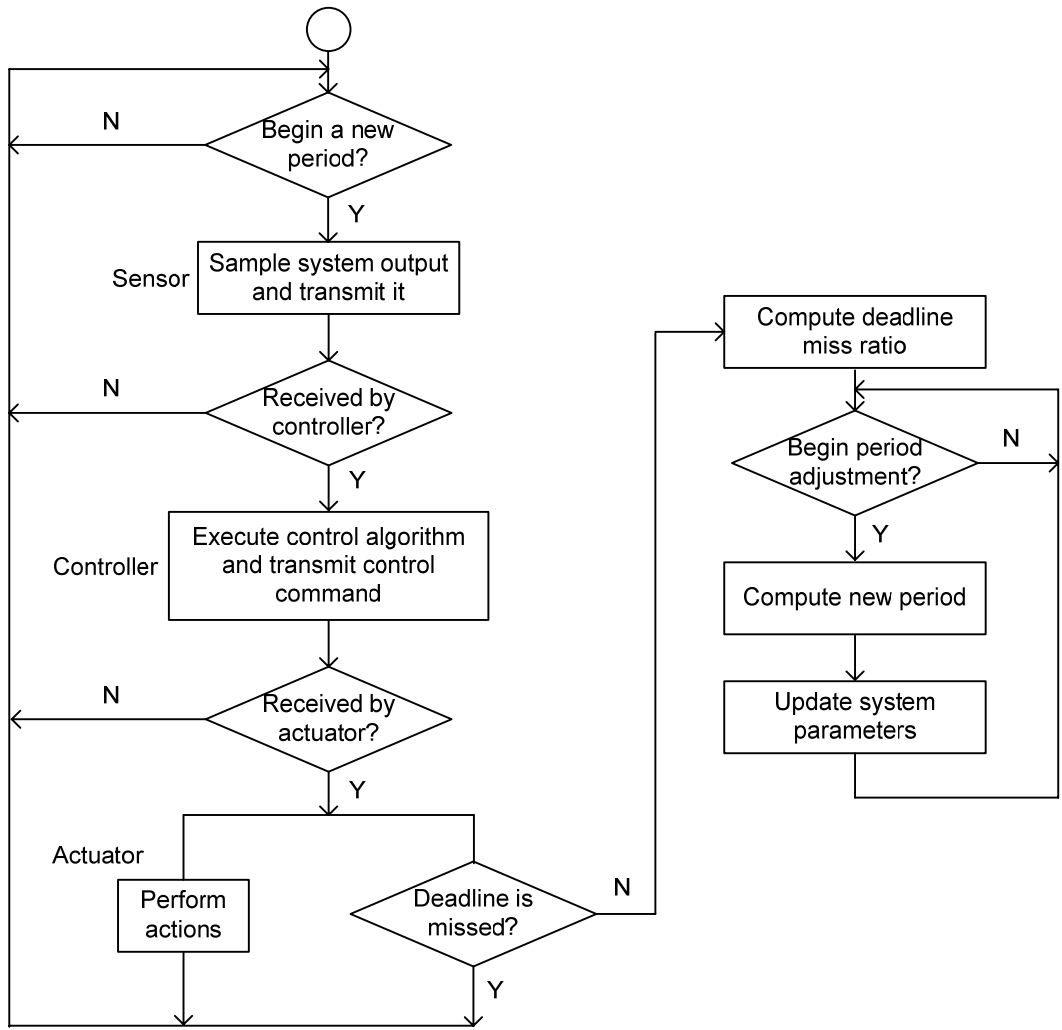

**Figure 3.** Flexible time-triggered sampling.

As shown in Figure 3, the flexible time-triggered sampling scheme operates as follows. Within every invocation interval, all nodes act almost the same as under time-triggered sampling, except for an additional module in the actuator. When the actuator receives a control command, it will not only perform actions on the physical process according to the control command but also judge whether or not this control command misses its deadline. For this purpose, the deadline of the control command will be issued by the smart sensor (when the sampled data is collected) and encapsulated in both data packets for sampled data and control command. If the deadline is not missed, the actuator will then report this to the sensor by directly sending it an arbitrary data. This information will be used by the smart sensor to compute the deadline miss ratio at the beginning of each invocation interval, i.e. each time the system starts to adjust the sampling period. The sensor will compute a new sampling period based on the observed deadline miss ratio every $T_{SPA}$ time units. The algorithm used will be given in Subsection 4.1. After the new sampling period is produced, the sensor will update the relevant internal parameter(s) accordingly.

Since sampling period variations will degrade QoC if fixed controller parameters are used in the controller, the controller parameters should be updated with respect to current sampling period. In practice, this can be achieved in two ways. The first way is that the sensor transmits the new sampling period (via the wireless network) to the controller using a separate data packet once a new sampling period is calculated. An acknowledgement will be sent by the controller if it successfully receives the sampling period data packet. The sensor will re-transmit this data packet after waiting for some specific time until an acknowledgement from the controller is received. Upon receiving a new sampling period, the controller will update the relevant controller parameters accordingly. In the second way, the current sampling period will always be encapsulated in the data packet for sampled data, which is sent from the sensor to the controller at the beginning of every period. The controller treats sampling period as an additional input variable and computes the control command with respect to both the sampled value of system output and the current sampling period. Using either of these ways, the variations in sampling period can be compensated for, at the expense of a slight increase in both computation and communication overheads. In this paper, the second method is adopted.

*4.1. Sampling Period Adaptation Algorithm*

As mentioned above, this paper uses feedback control theory to determine a new sampling period. Generally speaking, many control algorithms/techniques can be used in this context. In particular, the PID (proportional-integral-derivative) control algorithm, which is the most popular controller in practical control applications, is employed in this paper. Some reasons for the use of PID are explained as follows. Firstly, as a combination of three components, i.e., the proportional, integral and derivative components, the PID control algorithm has proved very effective in most control applications. Secondly, a PID controller can perform well even when the system model is unavailable, which is the case for many practical systems as well as the system considered in this paper, given that the controller coefficients are well tuned. Thirdly, the PID control algorithm is very simple, thereby inducing only a small computational overhead. This makes it easy to meet the requirements stemming from the limitations on the data processing capacities of smart sensors.

From a control perspective, the purpose of adjusting sampling period is to maintain the deadline miss ratio at a desired level. Let $\rho_r$ and $\rho(j)$ be the desired and measured deadline miss ratio, respectively, where $j$ corresponds to the $j$-th invocation of this algorithm. The sampling period is computed by:

$$\Delta h(j) = K_P(e(j) - e(j-1)) + K_I e(j) + K_D(e(j) - 2e(j-1) + e(j-2))$$
$$h(j) = \min\{h(j-1) - \Delta h(j), h_{\max}\} \quad (1)$$

where $K_P$, $K_I$, and $K_D$ are the proportional, integral, and derivative coefficients, respectively, $e(j)$ is the deadline miss ratio control error, and $h_{max}$ is the maximum allowable sampling period. Due to the unavailability of a mathematical model that describes explicitly the relationship between deadline miss ratio and sampling period, the coefficients $K_P$, $K_I$, and $K_D$ in (1) will be determined based on simulations in this paper. In general cases, $e(j)$ can be simply calculated as $e(j) = \rho_r - \rho(j)$. However, due to the inherent non-determinism of wireless communication, the measured deadline miss ratio may vary randomly from one invocation interval to another, even in the same network condition. To reduce the effect of this uncertainty as well as measurement noise, a low-pass filter is used in this paper when calculating $e(j)$, as given by:

$$e(j) = \rho_r - (\lambda\rho(j) + (1-\lambda)\rho(j-1)) \qquad (2)$$

where $\lambda$ is a forgetting factor that satisfies $0<\lambda\leq1$. A $\lambda$ close to 0 gives a smooth but slow estimate of the actual deadline miss ratio. The general case without the low-pass filter can be viewed as a special case of (2) where $\lambda$ is set to 1.

## 5. Performance Evaluation

In this section, simulations are conducted based on Matlab/TrueTime [32] to evaluate the performance of the proposed sampling scheme (denoted FTT), in comparison with the conventional non-adaptive time-triggered sampling scheme (denoted TT). For simplicity, suppose all control loops in the WCS have the same settings. The controlled process is a DC motor modeled by:

$$G(s) = \frac{1}{0.5s^2 + 6s + 10}$$

The DC motor is a physical component widely used in control systems. Details on its modeling can be found in [33]. The controller (in the control loop) for the DC motor uses the PID control algorithm, implemented as follows [3]:

**Procedure** PID controller {
  **Input**: $r(k)$, $y(k)$, $h$
  //$r(k)$: reference input (desired system output) at $k$-th sampling instant
  //$y(k)$: measured system output at $k$-th sampling instant
  $err(k) = r(k) - y(k)$;
  $P(k) = 100*err(k)$;
  $I(k) = I(k-1) + 200*h*(err(k) + err(k-1))/2$;
  $D(k) = 2*(err(k) - err(k-1))/h$;
  $u(k) = P(k) + I(k) + D(k)$;
  **Output**: $u$
  //$u(k)$: control command corresponding to the $k$-th sampling
}

It is worth noting that the above program is used in each controller within the control loops, which should not be confused with the sampling period adaptation module (see Section 4.1) in each sensor, although both of them use the PID control algorithm. Both the controlled process and the controller design are kept as common as possible to reflect the wide applicability of the proposed approach. The default sampling period is 10 ms, and the maximum allowable sampling period is $h_{max}$ = 30 ms. The reference input follows a square wave with a period of 4s. The data rate of ZigBee is 250 Kbps. The sizes of all data packets are 32 bytes. Since the flexible time-triggered sampling scheme is implemented in each control loop separately, the parameters $K_P$, $K_I$, and $K_D$ in (1) can be different from one loop to another. For simplicity, the same parameters are used in all sensors: $K_P$ = 0.007, $K_I$ = 0.006, $K_D$ = 0.003, $\rho_r$ = 10%, $\lambda$ = 0.7, and $T_{SPA}$ = 500 ms.

In this work, two typical scenarios featuring workload variations are examined, respectively, i.e., system reconfiguration and radio interference. While the bandwidth demands of all control loops can be regulated through sampling period adaptation, the bandwidth demand of an interfering node cannot be intentionally changed by the system.

*5.1. Scenario I: System Reconfiguration*

In the first set of simulations, the workload variations induced by dynamic reconfiguration of the system, in particular, the addition and removal of control loops, are studied. The simulation pattern is as follows. At time t = 0, two control loops, say loops 1 and 2, are active. Loops 3 and 4 are activated at t = 6s and deactivated at t = 12s, simultaneously.

The control performance of the four control loops is shown in Figure 4. Before loops 3 and 4 are activated, i.e., during time interval t = 0-6s, both control loops 1 and 2 achieve good performance with conventional time-triggered sampling, as can be seen from Figure 4(a). However, all control loops become unstable after the number of active control loops increases from 2 to 4 at time t = 6s, which causes the available bandwidth to be insufficient. In contrast, when the proposed flexible time-triggered sampling scheme is used, all control loops in the system remain stable and perform satisfactorily all the time, as shown in Figure 4(b).

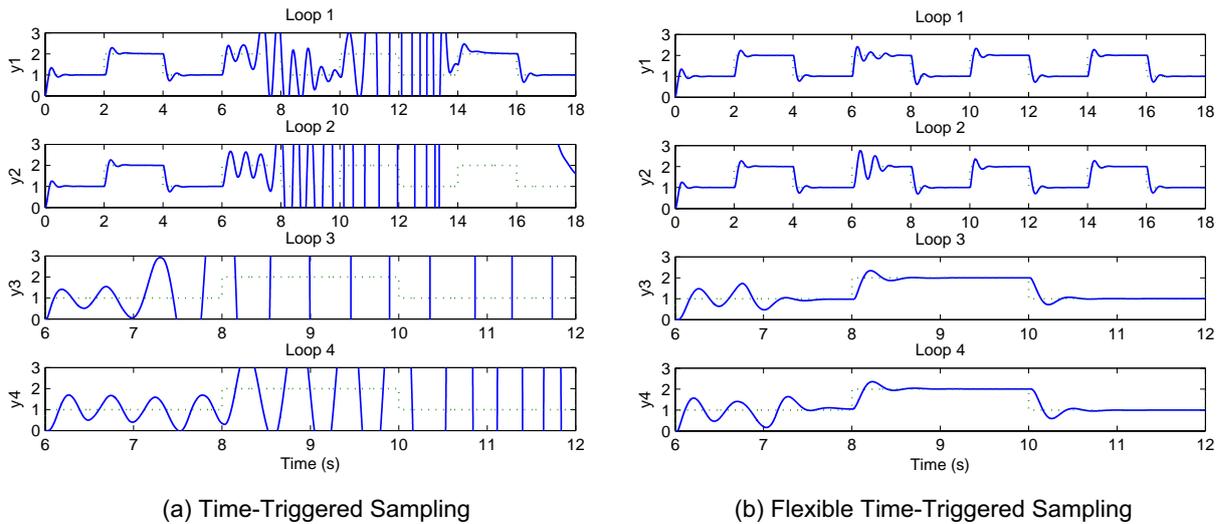

(a) Time-Triggered Sampling  (b) Flexible Time-Triggered Sampling

**Figure 4.** System output under system reconfiguration.

The sampling periods used in four smart sensors are depicted in Figure 5. It is apparent that the proposed flexible time-triggered sampling scheme dynamically adjusts the sampling period of each sensor at runtime, which is in contrast to the time-triggered sampling scheme that uses fixed periods.

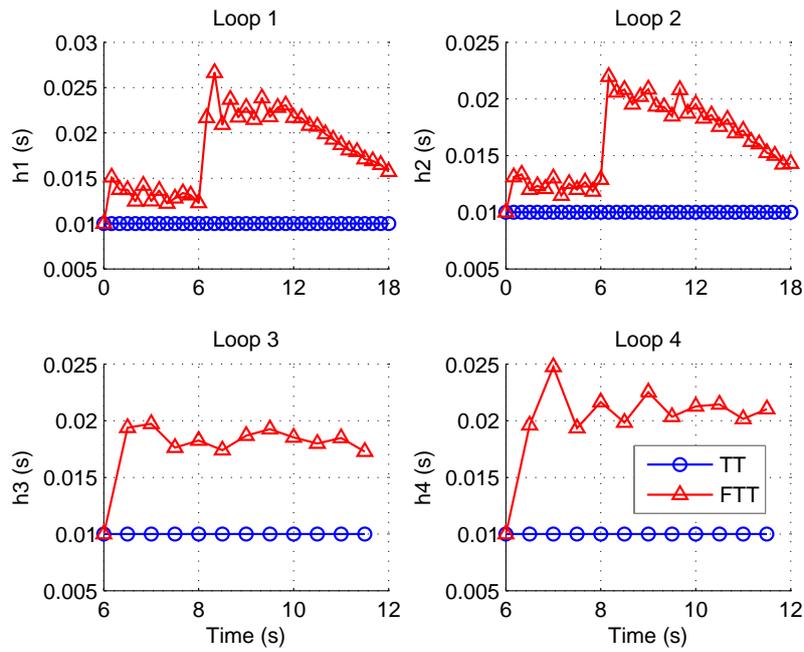

**Figure 5.** Sampling period under system reconfiguration.

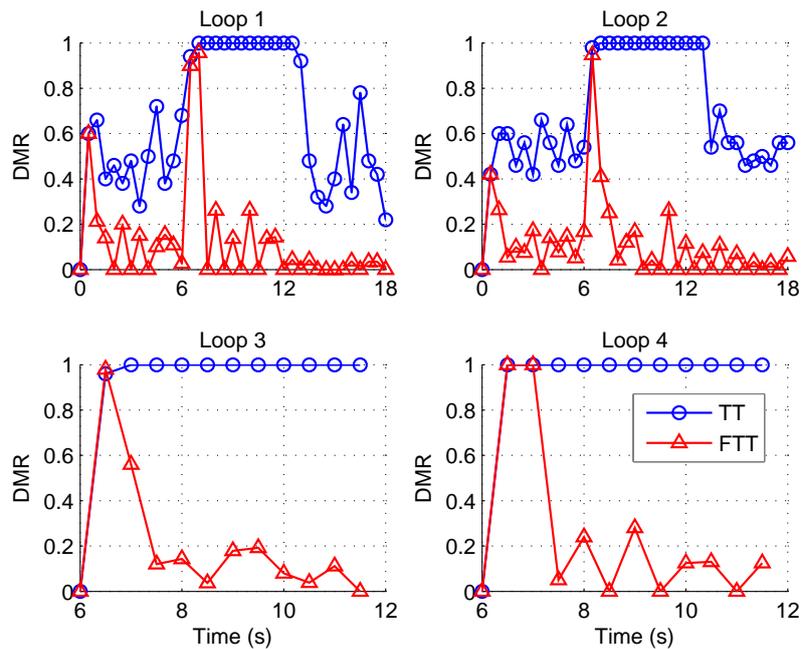

**Figure 6.** Deadline miss ratio under system reconfiguration.

The deadline miss ratios in control loops explain the difference in their control performance under different sampling schemes. As shown in Figure 6, all control loops suffer much severer deadline miss under time-triggered sampling than under flexible time-triggered sampling almost all the time. In particular, with time-triggered sampling, (almost) all control commands in the four control loops miss their deadlines during the time interval t = 6-12s, which yields system instability as shown in Figure 4(a). Under flexible time-triggered sampling, the deadline miss ratios in all control loops are well controlled and keep around the desired level most of the time. The variations in workload only incur some transient processes.

*5.2. Scenario II: Radio Interference*

The second set of simulations considers the impact of interfering radios. There are two (active) control loops in the system. At time t = 6s, two interfering nodes start to transmit data packets to another two nodes, respectively, and these transmissions last 6 seconds.

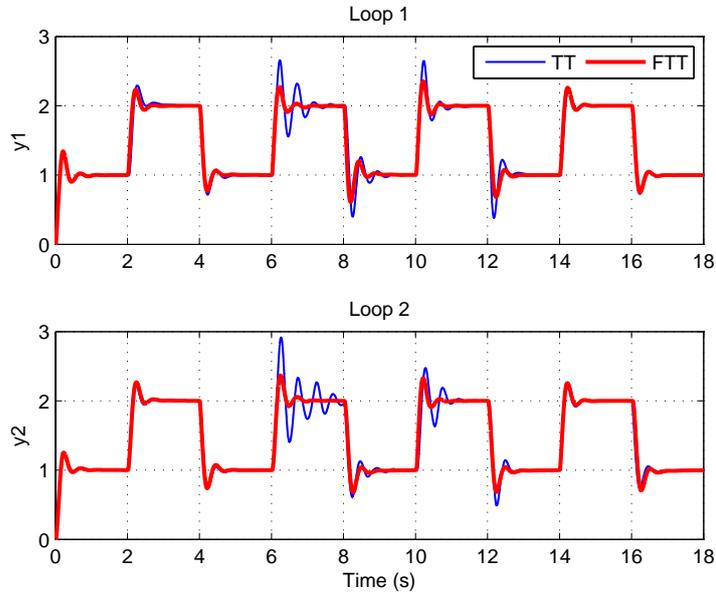

**Figure 7.** System output under slight interference.

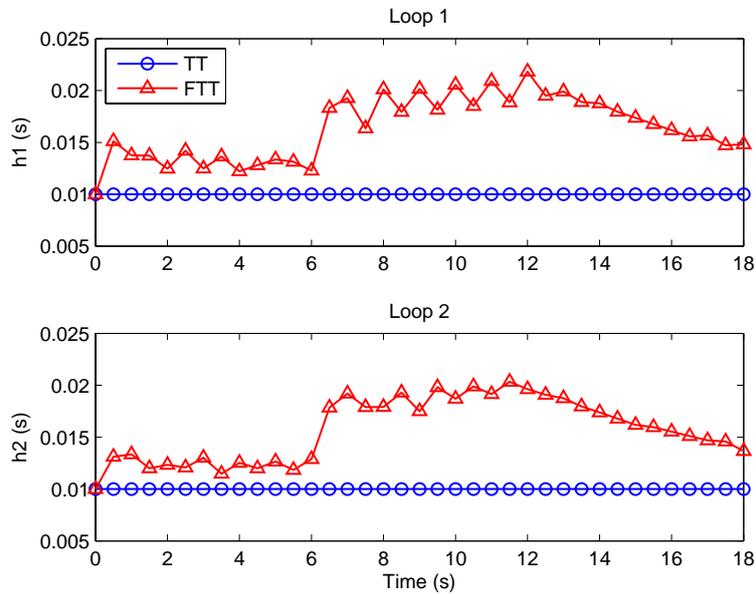

**Figure 8.** Sampling period under slight interference.

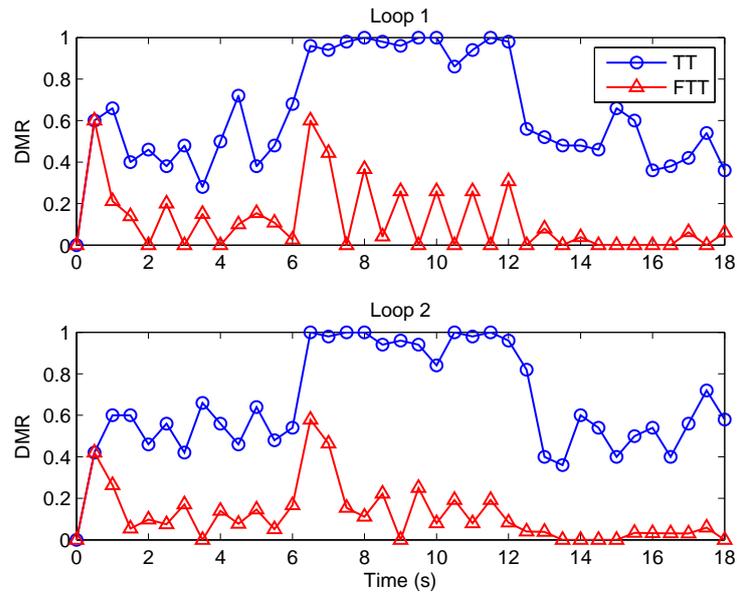

**Figure 9.** Deadline miss ratio under slight interference.

Figure 7 shows the control performance when each interfering node sends a packet every 10ms. It can be seen that, while the interference doesn't cause stability problems when the time-triggered sampling scheme is used, the proposed flexible time-triggered sampling scheme yields better control performance in both control loops (particularly) when the interference is present. Just as in Scenario I, this improvement mainly benefits from the dynamic adjustment of sampling periods in smart sensors, as shown in Figure 8. The deadline miss ratios in both control loops are also well controlled under flexible time-triggered sampling, see Figure 9. The lower deadline miss ratios achieved under flexible time-triggered sampling relative to time-triggered sampling explain the performance improvement shown in Figure 7.

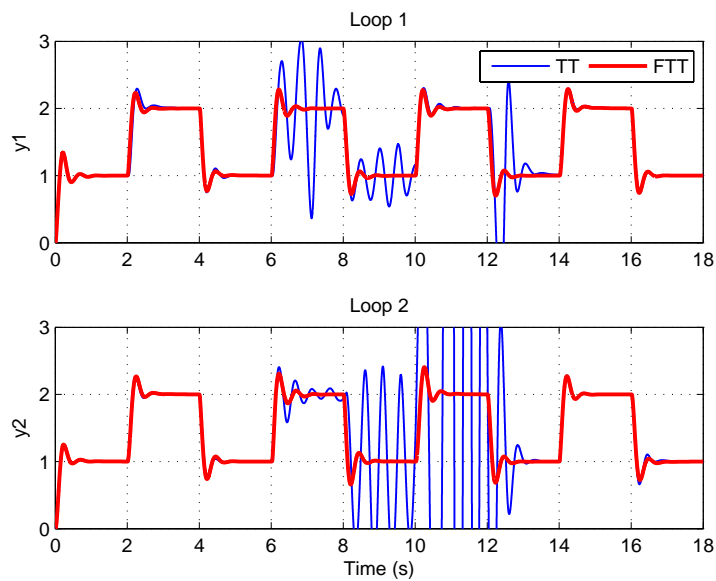

**Figure 10.** System output under severe interference.

Figure 10 shows the sharp difference in control performance with different sampling schemes when each interfering node sends a packet every 8ms. Under time-triggered sampling, the severe interference incurs system instability in both control loops. This doesn't happen when the flexible time-triggered sampling scheme is employed. The control performance of both loops remains good throughout the simulation. Again, this is a result of the sampling period adaptation that maintains the deadline miss ratios at relatively low levels (around the desired level). The graphs for sampling periods and deadline miss ratios in this case are omitted for brevity.

## 6. Conclusion

This paper has presented a flexible time-triggered sampling scheme for smart sensors that are used in WCSs. Based on time-triggered sampling, this scheme enhances the flexibility and resource efficiency of the system through adapting the sampling period at runtime. Feedback control technology is used to determine the sampling period that attempts to maintain the deadline miss ratio in each control loop at a desired level. Extensive simulations have been conducted to evaluate the performance of the proposed scheme. From the simulation results, it can be argued that the proposed sampling scheme is able to deal with dynamic and unpredictable variations in workload induced by e.g. system reconfiguration and radio interference, while providing QoC guarantees. This makes it well suited for smart sensor-based WCSs that operate in dynamic environments.

Our future work in this direction includes: 1) development of an experimental WCS based on smart wireless sensors to further validate the proposed approach; 2) applications of advanced control techniques (e.g. fuzzy control) in the sampling period adaptation module.

## References and Notes

1. Willig, A.; Matheus, K.; Wolisz, A. Wireless technology in industrial networks. *Proceedings of the IEEE* **2005**, *93*(6), 1130-1151.
2. Mathiesen, M.; Thonet, G.; Aakwaag, N. Wireless ad-hoc networks for industrial automation: current trends and future prospects. In Proceedings of the IFAC World Congress, Prague, Czech Republic, Jul 3-8, 2005; Piztek, P., Ed.; Elsevier: Amsterdam, The Netherlands, 2006.
3. Xia, F. Feedback scheduling of real-time control systems with resource constraints. Ph.D. Thesis, Zhejiang University, 2006.
4. Pellegrini, F.D.; Miorandi, D.; Vitturi, S.; Zanella, A. On the Use of Wireless Networks at Low Level of Factory Automation Systems. *IEEE Trans. on Industrial Informatics* **2007**, *2*(2), 129-143.
5. Xia, F.; Tian, Y.-C.; Li, Y.; Sun, Y. Wireless Sensor/Actuator Network Design for Mobile Control Applications. *Sensors* **2007**, *7*, 2157-2173.
6. Steigmann, R.; Endresen, J. Introduction to WISA, White Paper, V2.0, ABB, July 2006.
7. Ploplys, N.; Kawka, P.; Alleyne, A. Closed-Loop Control over Wireless Networks. *IEEE Control Systems Magazine* **2004**, *24*(3), 58-71.
8. Astrom, K.J.; Bernhardsson, B. Comparison of periodic and event based sampling for first-order stochastic systems. In Proceedings of the 14th World Congress of IFAC, Beijing, P.R. China, Jul 5-9, 1999; Chen, H.-F., Cheng, D.-Z., Zhang, J.-F., Eds.; Elsevier: Amsterdam, The Netherlands, 1999.

9. Otanez, P.; Moyne, J.; Tilbury, D. Using deadbands to reduce communication in networked control systems. *Proc. of American Control Conference* **2002**, *4*, 3015-3020.
10. Miskowicz, M. Send-on-delta concept: an event-based data reporting strategy. *Sensors* **2006**, *6*, 49-63.
11. Suh, Y.S. Send-On-Delta Sensor Data Transmission with a Linear Predictor. *Sensors* **2007**, *7*, 537-547.
12. Miskowicz, M. Asymptotic Effectiveness of the Event-Based Sampling according to the Integral Criterion. *Sensors* **2007**, *7*, 16-37.
13. Miskowicz, M. The Event-Triggered Sampling Optimization Criterion for Distributed Networked Monitoring and Control Systems. In Proceedings of International Conference on Industrial Technology (ICIT'03), Maribor, Slovenia, Dec 10-12, 2003; pp. 1083-1088.
14. Astrom, K. J. Event based control. In *Analysis and Design of Nonlinear Control Systems*, Astolfi, A., Marconi, L., Eds.; Springer-Verlag: Heidelberg, Germany, 2007.
15. Wang, X.; Wang, S.; Ma, J. An Improved Particle Filter for Target Tracking in Sensor Systems. *Sensors* **2007**, *7*, 144-156.
16. Ramamurthy, H.; Prabhu, B. S.; Gadh, R.; Madni, A. Wireless Industrial Monitoring and Control Using a Smart Sensor Platform. *IEEE Sensors Journal* **2007**, *7*(5), 611-618.
17. Willett, R.; Martin, A.; Nowak, R. Backcasting: Adaptive Sampling for Sensor Networks. In Proceedings of the International Symposium on Information Processing in Sensor Networks, 2004; pp. 124-33.
18. Nguyen, V.H.; Suh, Y.S. Improving Estimation Performance in Networked Control Systems Applying the Send-on-delta Transmission Method. *Sensors* **2007**, *7*, 2128-2138.
19. Cervin, A.; Eker, J.; Bernhardsson, B.; Årzén, K.-E. Feedback-Feedforward Scheduling of Control Tasks. *Real-Time Systems* **2002**, *23*(1), 25-53.
20. Martí, P.; Lin, C.; Brandt, S.; Velasco, M.; Fuertes, J.M. Optimal State Feedback Based Resource Allocation for Resource-Constrained Control Tasks. In Proceedings of the 25th IEEE Real-Time Systems Symposium (RTSS04), Lisbon, Portugal, 2004; pp. 161-172.
21. Xia, F.; Li, S.; Sun, Y. Neural Network Based Feedback Scheduler for Networked Control System with Flexible Workload. *Lecture Notes in Computer Science* **2005**, *3611*, 237-246.
22. Xia, F.; Liu, L.; Sun, Y. Flexible Quality-of-Control Management in Embedded Systems Using Fuzzy Feedback Scheduling. *Lecture Notes in Artificial Intelligence* **2005**, *3642*, 624-633.
23. Xia, F.; Shen, X.; Liu, L.; Wang, Z.; Sun, Y. Fuzzy Logic Based Feedback Scheduler for Embedded Control Systems. *Lecture Notes in Computer Science* **2005**, *3645*, 453-462.
24. Xia, F.; Sun, Y. Control-Scheduling Codesign: A Perspective on Integrating Control and Computing. In *Dynamics of Continuous, Discrete and Impulsive Systems - Series B: Applications and Algorithms, Special Issue on ICSCA'06*, 2006; pp. 1352-1358.
25. Li, Z.; Chow, M.-Y. Adaptive Multiple Sampling Rate Scheduling of Real-time Networked Supervisory Control System - Part II. In *Proc. IEEE IECON'06* **2006**, 4610-4615.
26. Kawka, P. A.; Alleyne, A. G. Stability and Feedback Control of Wireless Network Systems. In Proceedings of the American Control Conference (ACC05), Portland, OR, 2005; Volume 4, pp. 2953-2959.
27. Colandairaj, J.; Irwin, G. W.; Scanlon, W. G. Wireless networked control systems with QoS-based sampling. *IET Control Theory & Applications* **2007**, *1*(1), 430-438.

28. Xia, F.; Liu, L.; Li, S.; Sun, Y. Integrated Feedback Scheduling of Networked Control Systems. In *Dynamics of Continuous, Discrete and Impulsive Systems - Series B: Applications and Algorithms, Special Issue on ICSCA'06*, 2006; pp. 3274-3280.
29. Almeida, L.; Pedreiras, P.; Fonseca, J. A. The FTT-CAN protocol: Why and how. *IEEE Transactions on Industrial Electronics* **2002**, *49*(6), 1189-1201.
30. Pedreiras, P.; Gai, P.; Almeida, L.; Buttazzo, G. FTT-Ethernet: A Flexible Real-Time Communication Protocol That Supports Dynamic QoS Management on Ethernet-Based Systems. *IEEE Transactions on Industrial Informatics* **2005**, *1*(3), 162-172.
31. http://www.zigbee.org, accessed Oct 1, 2007.
32. Andersson, M.; Henriksson, D.; Cervin, A.; Arzen, K.-E. Simulation of Wireless Networked Control Systems. In Proceedings of the 44th IEEE CDC and ECC, Seville, Spain, 2005; pp. 476-481.
33. http://www.library.cmu.edu/ctms/, accessed Oct 1, 2007.